\documentclass[%
 reprint,
superscriptaddress,
showpacs,preprintnumbers,
 amsmath,amssymb,
 aps,
]{revtex4-1}

\usepackage{graphicx}
\usepackage{dcolumn}
\usepackage{bm}
\usepackage{color}
\renewcommand{\thefigure}{\arabic{figure}}
\def\be{\begin{equation}}
\def\ee{\end{equation}}

\begin{document}
\renewcommand{\thefigure}{\arabic{figure}}

\title{ Dirac fermion and superconductivity in two-dimensional transition-metal MOH (M= Zr, Hf)}
\author{Ali Ebrahimian}
\email{ebrahimian.Al@fs.lu.ac.ir}
\email{aliebrahimian@ipm.ir}
\affiliation{Department of Physics, Lorestan University, Khoramabad, Iran}
\affiliation{School of Nano Science, Institute for Research in Fundamental Sciences (IPM), Tehran 19395-5531, Iran}
\author{Mehrdad Dadsetani}
\affiliation{Department of Physics, Lorestan University, Khoramabad, Iran}
\author{Reza Asgari}
\affiliation{School of Nano Science, Institute for Research in Fundamental Sciences (IPM), Tehran 19395-5531, Iran}
\affiliation{School of Physics, Institute for Research in Fundamental Sciences (IPM), Tehran 19395-5531, Iran}
\date{\today}

\begin{abstract}
Discovery of the new two-dimensional (2D) Dirac semimetals incorporating both superconductivity and the topological band structure has provided a novel platform for realizing the intriguing applications of Dirac fermions and Majorana quasiparticles, ranging from high-speed quantum devices at the nanoscale to topological quantum computations. In this work, utilizing first-principles calculations and symmetry analysis, we introduce MOH (M= Zr, Hf) as a new topological superconductor with Dirac points close to a Fermi level which are connected with nearly flat edge states as a striking feature of topological semimetals. Our calculations show that ZrOH as a 2D topological semimetal can exhibit superconductivity and is a novel platform for studying the interplay between superconductivity and Dirac states in low-dimensional materials.
\end{abstract}

\maketitle

\section{Introduction}
Nowadays, the boundary between different branches of science is fading and novel ideas developed by researchers in one area have found broad applications in others. Condensed matter science and other related subjects have been a source of many discoveries, some examples would be $^{3}$He-B and $^{3}$He-A phases as topological superfluid, graphene, Weyl semimetal and topological order. After introducing the Dirac (Weyl) superconductor (SC)~\cite{1}, search for topological superconducting materials has become one of the main topics in modern physics. Up to now, most of the reported topological superconductors belong to two categories including insulator/SC heterostructure~\cite{2, 3, 4} and doped topological insulator~\cite{5, 6}. However, the experimental observation of superconductivity in these materials is very challenging~\cite{7, 8}.
Recent theoretical studies show that the unique orbital texture of Dirac points in doped Dirac semimetals allows odd-parity Cooper pairing between electrons with different quantum numbers~\cite{9, 10}. In Dirac semimetals, the energy bands are doubly degenerate with the conduction bands intersecting with the valence bands at Dirac points, such that the band-crossing points are fourfold degenerate and their low-energy excitations are linearly dispersing Dirac fermions~\cite{11}.
This degeneracy in general is not topologically protected~\cite{12} and requires the protection of the special space group symmetries in which the band-crossing remains intact as symmetry-protected degeneracies~\cite{13}.

Two-dimensional (2D) materials with tunable topological and electronic properties via surface termination, are a paradigm for designing materials that incorporate both superconductivity and the topological band structure~\cite{14, 15, 16}. More speculative effects arise in such  systems providing platforms to exploit novel properties in spintronics and topological qubit. 

Here, we report an investigation of the influence of element substitution in transition metal halide MX (M=Zr, Hf; X=Cl, Br)
monolayers on their topological nature and propose a method for creating new topological SC from these monolayers, based on the quantitative first-principles calculations and symmetry analysis.
Layered compounds ZrCl and ZrBr, consisting of tightly bound double hexagonal Zr atomic layers sandwiched between two halogen atomic layers, have been synthesized in experiment~\cite{17, 18, 19}. More excitingly, L. Zhou et al~\cite{20} predicted that MX monolayers constitute a novel family of robust Quantum spin Hall (QSH) insulator. In this work, we predict that by replacing halogen atoms of these monolayers with Hydroxyl (OH) groups, the conduction (valence) bands shift down (up) and the band inversion happens at the $\Gamma$ point leading to the appearance of the Dirac cone and high density of states (DOS) at the Fermi energy. Moreover, during this change of the element, the Zr-Zr (Hf-Hf) distance decreases and the highly disperse and flat bands appear in the vicinity of the Fermi level providing the necessary condition for high superconducting transition temperature ($T_{\rm c}$)~\cite{21, 22, 23}. In other words, our calculations show that ZrOH (HfOH) is a topological SC with a superconducting $T_{\rm c}$ of greater than 20 K in addition to Dirac cone at 0.05 eV above the Fermi level which makes it an ideal platform to study Dirac (Weyl) physics and Majorana fermion quasiparticle.

\begin{figure}[b]
	\centering
	\includegraphics[width=1.0\linewidth]{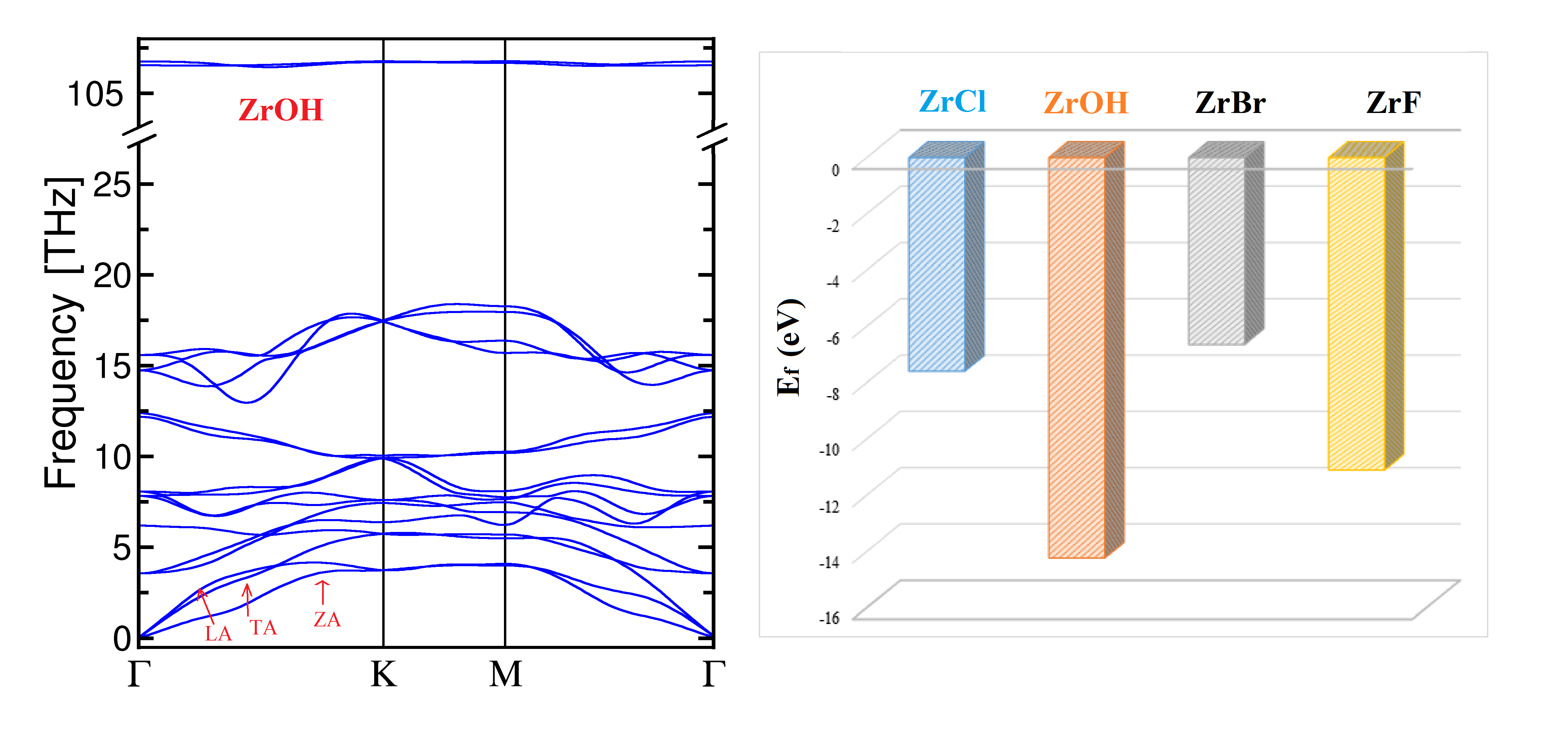}\\
	\caption{(Color online) Left panel: The phonon dispersions of optimized ZrOH nanosheets. The sound velocities in the $\Gamma$-K direction are obtained as 1.55 and 1.48 km/s for longitudinal and transverse atomic motions respectively. Right panel: The formation energies of surface groups for single-layer ZrX (X= OH, Cl, Br, F).
		\label{fig:mass}}
\end{figure}

\begin{figure*}[t]
\centering
\includegraphics[width=\textwidth]{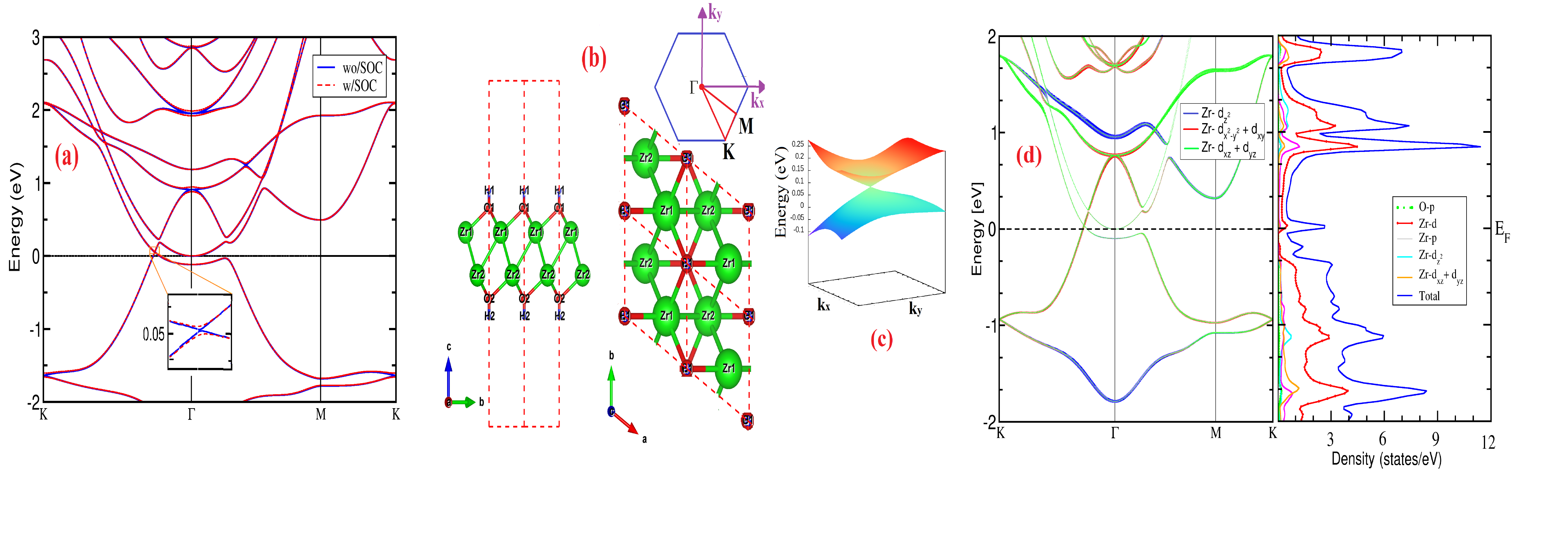} \\
\caption{(Color online) (a) The calculated bulk band structure of ZrOH using HSE06. The enlarged plot at the Dirac point with/without SOC shows the linear band dispersion around crossing point. (b) Side view (left), top view (lower right) of the crystal structure of ZrOH and the first BZ of ZrOH (upper right). (c) The 3D plot of the Dirac cone. (d) Left: The orbital-projected band structures of ZrOH shows that the band near the Fermi level is mostly composed of Zr-d orbitals. Right: DOS of ZrOH indicates a large density of
states at the Fermi level.
\label{fig:E}}
\end{figure*}

This paper is organized as follows. In Sec. II, we briefly introduce the theory and simulation methods. Sec. III is devoted to the numerical results of the study, focusing on the electronic and topological as well as electron-phonon coupling properties of the hexagonal ZrOH. Last but not least, we summarize our results in Sec. IV.

\section{Theory and simulation methods}

Our calculations are based on density functional theory~\cite{24, 25, 26, 27, 28, 29, 30, 31, 32, 33}. 
The electronic structure simulations are calculated using density functional theory which is implemented in WIEN2k code~\cite{24} as well as the FHI-aims code package~\cite{25} to meet high accuracy requirements. To avoid the possible underestimation of the band gap within generalized gradient
approximation (GGA), in addition to PBE-type~\cite{26} of generalized gradient approximation functional, the hybrid density functional (HSE06)~\cite{27, 28} and the mBJ potential~\cite{29} are used in the calculations. A 20$\times$20$\times$1 Monskhorstpack k-point mesh was used in the computations. The spin-orbit coupling (SOC) is included consistently within the second variational method. A 21\AA~ thick vacuum layer is used to avoid interactions between nearest layers. The WannierTools code~\cite{30} was used to investigate the topological properties based on maximal localized functions tight-binding model ~\cite{31} that was constructed by using the Wannier90 package~\cite{32} with Zr (Hf) d orbitals as projectors. The surface state spectra are calculated using the iterative Green's function method~\cite{30, 33}.

The calculations of the electron-phonon coupling and superconductivity are based on the density functional perturbation theory (DFPT)~\cite{34}. Quantum Espresso~\cite{35} packages are employed with Ultrasoft Pseudopotentials in the phonon calculations. The energy cutoff is 50 Ry (500 Ry) for wave function (charge densities) calculations. 48$\times$48$\times$1 k grid and 6$\times$6$\times$1 q grid are used in DFPT calculations. The electron-phonon coupling strength is given by

\be\label{dynamical_term}
\lambda_{\rm q\nu}=\frac{1}{\pi N_{\rm F}}\frac{\pi^{\prime\prime} _{\rm q\nu}}{\omega^2_{\rm q \nu}  }
\ee
where N$_{\rm F}$ is the density of states at the Fermi level and $\omega_{\rm q \nu}$ is a phonon frequency of mode $\nu$ at wave vector q. The electron-phonon quasiparticle line width $\pi^{\prime\prime} _{\rm q\nu}$ can be written as

\be\label{dynamical_term}
\pi^{\prime\prime} _{\rm q\nu}=\pi \omega_{q \nu} \sum_{\rm mn,k} \mid g^{\nu}_{\rm mn}(k,q) \mid^2 \delta(\epsilon_{\rm nk}) \delta(\epsilon_{\rm mk+q})
\ee
where $\epsilon_{\rm nk}$ is the energy of the KS orbital and the dynamical matrix $g^{\nu}_{\rm mn}(k,q)$ reads

\be\label{dynamical_term}
g^{\nu}_{\rm mn}(k,q)=(\frac{\hbar}{2M \omega_{\rm q \nu}})^{1/2} \langle \Psi_{\rm kn}  \mid \frac{\mathrm{d}V_{\rm scf}}{\mathrm{d}u_{\rm q \nu}}\cdot e_{\rm q \nu} \mid \Psi_{\rm k+qm} \rangle
\ee

$M$ and $e_{\rm q \nu}$ represent the mass of the atom and the unit vector along $u_{\rm q \nu}$ respectively and $\frac{\mathrm{d}V_{\rm scf}}{\mathrm{d}u_{\rm q \nu}}$ denote the deformation potential at the small atomic displacement $\mathrm{d}u_{\rm q \nu}$ of the given phonon mode. The transition temperature $T_{\rm c}$ can be estimated by Allan-Dynes modified McMillan formula~\cite{36, 37}:

\be\label{dynamical_term}
T_{\rm c}=\frac{f_{\rm 1}f_{\rm 2}\omega_{\rm ln}}{1.2}exp[-\frac{1.04(1+\lambda)}{\lambda-\mu^*(1+0.62\lambda)}]
\ee
where $\mu^*$ is the Morel-Anderson Pseudopotential~\cite{37} and is usually set between 0.1 and 0.2. The parameters $f_{\rm 1}$ and $f_{\rm 2}$ are strong-coupling and shape correction respectively. $\lambda$ and $\omega_{\rm ln}$ is given by
\be\label{dynamical_term}
\lambda=\sum_{\rm q \nu}\lambda_{\rm q \nu}
\ee

\be\label{dynamical_term}
\omega_{\rm ln}=exp [ \frac{2}{\lambda}  \int\mathrm{d}{\bf \omega\, \mathrm{\frac{ln\omega}\omega}  \mathrm{\alpha^2F(\omega)}}]
\ee
\be\label{dynamical_term}
\mathrm {\alpha^2F(\omega)}= \frac{1}{2} \sum_{\rm  \nu} \int_{\rm BZ}\frac{\mathrm{d}{\bf q\ }}{\Omega_{\rm BZ}}{\mathrm{\lambda_{q\nu}\omega_{q \nu}} \delta(\omega-\mathrm{\omega_{q \nu}})}
\ee
It is worth mentioning that by setting $f_{\rm 1}=f_{\rm 2}=1$, Eq. 7 is reduced to a well-known Allan-Dysnes MacMillan formula which works in a case that $\lambda$ is smaller than unity. In our case, as we will discuss in the numerical results, the $\lambda$ is less than unity and therefore, $T_{\rm c}$  is the same obtained within those formulas.

\section{Results and discussions}

MOH (M= Zr, Hf) monolayer has a similar structure to MX monolayers~\cite{20} which is composed of tightly bound double hexagonal M atomic layers sandwiched between two hexagonal hydroxyl group (OH) layers in layering sequence of H-O-M-M-O-H. Compared to the MX sheets, the MOH layers prefer smaller bond lengths and lattice constants as the ground state structure. Due to similarities of electronic structures of ZrOH and HfOH, we focus on the ZrOH monolayer. The results for HfOH are given in Appendix C. Furthermore, the band structures calculated by mBJ are very similar to those obtained within the HSE06 potential in the vicinity of the Fermi energy, therefore, we mainly use HSE06 potential in our discussions.

The calculated phonon spectra (Fig.~\ref{fig:mass}) show no imaginary modes, implying the structural stability of the ZrOH monolayer. Note that the phonon dispersions are calculated by using the 2D Coulomb cut off within DFPT to eliminate the spurious long-range interactions with the periodic copies and correctly account for the long wavelength of polar-optical phonons in 2D framework~\cite{38, 39}. The calculated 2D linear elastic constants ($C_{\rm 11}=149.23$, $C_{\rm 12}=-26.7$, $C_{\rm 13}=41.22$, $C_{\rm33}=34.43$ and $C_{\rm 55}$=37.88 Nm$^{-1}$) satisfy the Born stability criteria~\cite{40} of hexagonal sheet, indicating robust mechanical stabilities for ZrOH.
Moreover, the formation energy of ZrOH (which is defined as $E_{\rm f}$=$E_{\rm tot}$(ZrOH)-$E_{\rm tot}$(Zr)-$E_{\rm tot}$(OH) where $E_{\rm tot}$(ZrOH) and $E_{\rm tot}$(Zr) stand for the total energies of ZrOH and Zr monolayer, respectively, and $E_{\rm tot}$(OH) is the total energy of O$_{\rm 2}$+H$_{\rm 2}$) is computed to consider its thermodynamic stability (Appendix A). The calculated formation energy of ZrOH is negative (Fig.~\ref{fig:mass}), implying exothermic functionalization of OH on the pristine Zr monolayer. Since the formation energy of ZrOH is larger than those of ZrX, probably its experimental production may be more feasible.

\begin{figure*}[!htp]
	\includegraphics[width=0.95\linewidth]{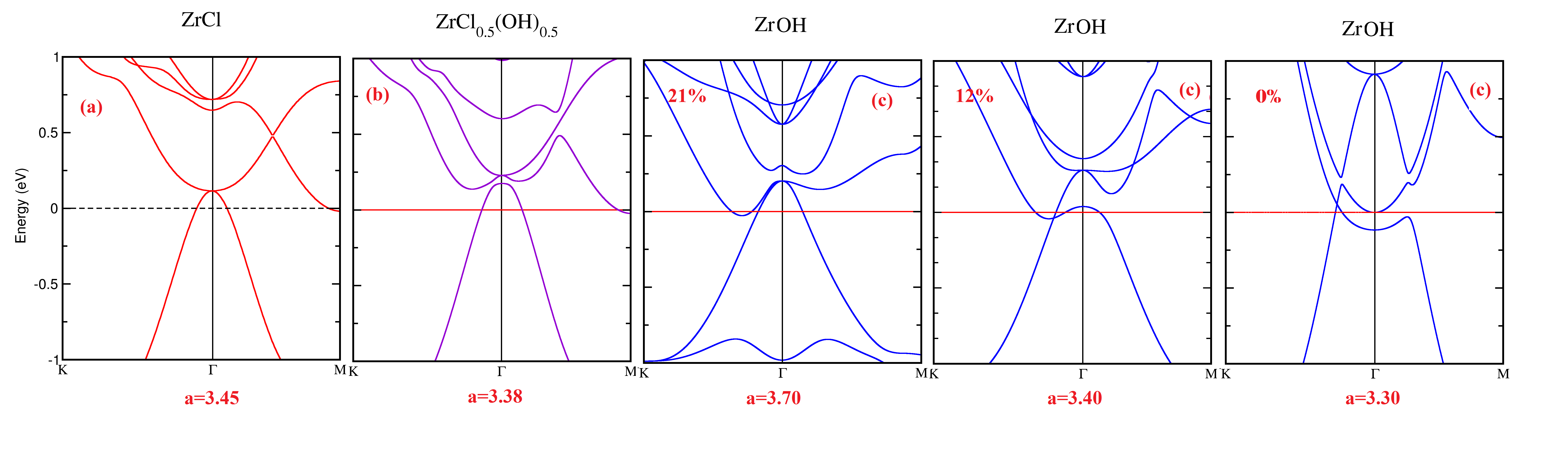}\\
	\caption{(Color online) The band structures of ZrCl (a), ZrCl$_{\rm 0.5}$(OH)$_{\rm 0.5}$ (b). (c) The evolution of closest bands to the Fermi level under lattice strain for ZrOH.
		\label{fig:hf}}
\end{figure*}

\begin{figure*}[t]
\centering
\includegraphics[width=0.88\textwidth]{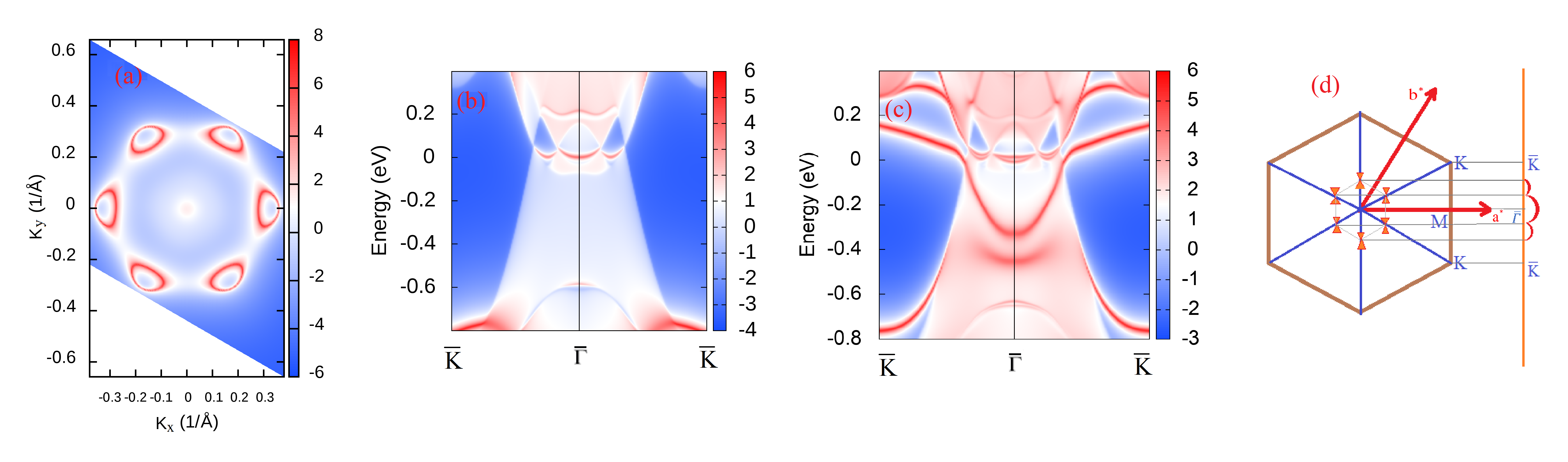}\\
\caption{(Color online) The calculated Fermi surface and (b and c) momentum-resolved edge density of states in the (1, 0) edge for ZrOH.  Edge states (red trace) connects the projected Dirac cones in the presence (c) and absence (b) of SOC. (d) A schematic figure of the formation of nearly flat edge states and the corresponding projected Dirac points in (1, 0) edge BZ.
\label{fig:sub}}
\end{figure*}

As shown in Fig.~\ref{fig:E}, the valence and conduction bands cross each other along the $\Gamma$-K making a Dirac point band degeneracy at 0.05 eV above Fermi level. The three-dimensional (3D) plot for the Dirac cone (Fig.~\ref{fig:E} (c)) presents a notably anisotropic character around the Dirac point which can be used for manipulating the propagation of carriers~\cite{41}. When taking into account the spin-orbit coupling (SOC) effects, the Dirac band structure is well retained except for a negligible band gap, about 2.5 meV, at the Dirac point (Fig.~\ref{fig:E} (a)). In other words, in the presence of the SOC the main feature of the band structure is unchanged.

Just as the transition metal halide ZrX, Zr-d orbitals dominate the band near the Fermi level (Fig.~\ref{fig:E}). In comparison to ZrX, OH termination pushes downward the conduction band beneath the Fermi level resulting in band crossing. Due to the D$_{\rm 3d}$ symmetry of the structure, the point group of the wave vector at the high-symmetry point $\Gamma$, K and M is D$_{\rm 3d}$, D$_{\rm 3}$, and C$_{\rm 2h}$ respectively. The D$_{\rm 3d}$ point group at the $\Gamma$ point splits the Zr-d orbitals into three groups (d$_{\rm xy}$, d$_{\rm x^2-y^2}$), (d$_{\rm xz}$, d$_{\rm yz}$) and d$_{\rm z^2}$. The point group along the high symmetry line $\Gamma$-K ($\Gamma$-M) is C$_{\rm 2}$ (C$_{\rm s}$) which has two irreducible representations. The calculated irreducible representations of the point group of the crossing bands show that these bands have different representations A and B along the $\Gamma$-K. So they cannot interact and mix with each other making a Dirac point near the Fermi level. Furthermore, since off-symmetry point has C$_{\rm 1}$ point group, the band crossing is avoided for every general nonsymmetric point. Therefore, the dispersion relation is linear at the vicinity of the Dirac point forming a Dirac cone around the Dirac point (Fig. ~\ref{fig:E} (c)). As the SOC is taken into account, it lifts the degeneracy of the points and opens a negligible gap approximately 2.5 meV (or 29 K). However, the SOC-induced gap can be neglected in cases with temperature greater than 29 K.

To investigate the reason of the band crossing, we change the lattice constant {\bf a} while analyzing the band order associated with atomic orbitals. Furthermore, considering the effects of substitution of Cl with OH group, we calculate the band structures of ZrCl and ZrCl$_{\rm 0.5}$OH$_{\rm 0.5}$ monolayers. As shown in Fig.~\ref{fig:hf} by replacing 50 percent of Cl atoms with OH groups, the lattice constant {\bf a} becomes smaller lowering the energy of the conduction bands above the $\Gamma$ point. When the Cl atoms are replaced completely by OH groups (ZrOH), the lattice constant {\bf a} reduces to 3.30 \AA, causing the conduction and valence band inversion at the $\Gamma$ point. In ZrOH, by increasing the lattice constant {\bf a}, the distance between Zr atoms increases so the interactions between Zr atoms and hybridization between Zr-d$_{\rm xy/x^2-y^2}$ (Zr-d$_{\rm Z^2}$ ) diminishes (enhances) which lowers (raises) the energy of the bands composed from Zr-d$_{\rm xy/x^2-y^2}$ (Zr-d$_{\rm Z^2}$ ) orbitals resembling the band structure of ZrCl monolayer.

Because of separation of the conduction and valence bands, the Fu-Kane formula of topological invariants can be applied to ZrOH and the $Z_{\rm 2}$ invariant is determined by examining the parity eigenvalues of the valence band at time reversal invariant momenta (TRIM) points~\cite{42}. The parity products for occupied states at two TRIM of $\Gamma$, M are calculated to be ''+'' and ''-'' respectively which gives $Z_{\rm 2}$=1, indicating the nontrivial topological phase. The calculated Fermi surface in Fig.~\ref{fig:sub} (a) shows that there are six equivalent Dirac points in Brillouin zone (BZ) which is consistent with C$_{\rm 3}$ rotation and time-reversal symmetry of the structure. The calculated  momentum-resolved surface density of states along the high symmetry line of $\bar{K}$-$\bar\Gamma$-$\bar{K}$  in the (1, 0) edge BZ (Fig.~\ref{fig:sub} (b and c)) shows the presence of edge modes (red line) connecting two projected Dirac points. It can be seen that without the SOC effect, a nearly flat edge state links the adjacent projected Dirac points. When the SOC is switched on, a SOC-induced spin splitting appears for the edge states. At the (1, 0) edge, because four bulk Dirac nodes project onto two points in the edge Brillouin zone, there are four Dirac nodes in this edge (Fig.~\ref{fig:sub} (d)).

\begin{figure}
\centering
\includegraphics[width=0.9\linewidth]{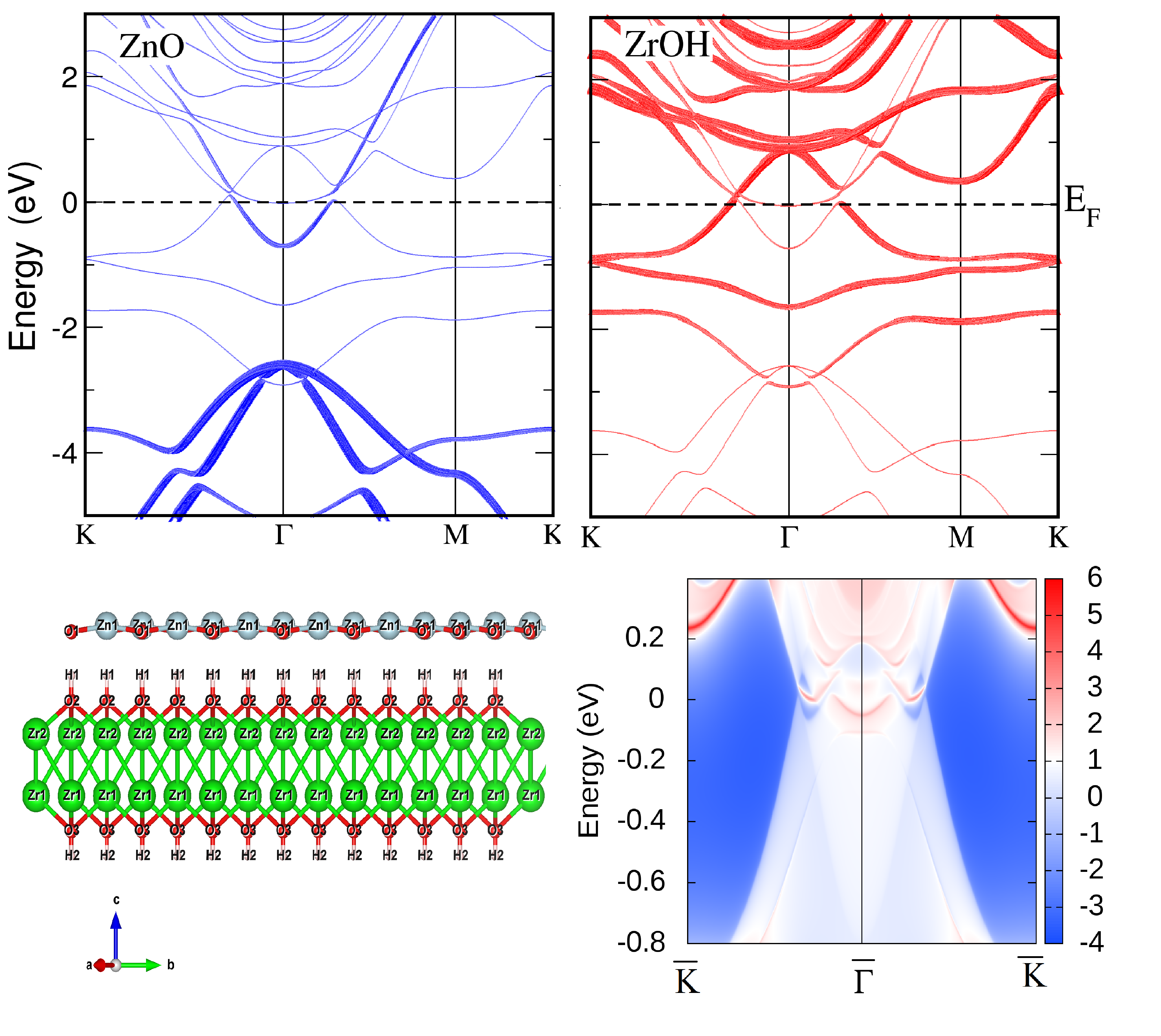}\\
\caption{(Color online) Upper panel: The calculated electronic band structures of ZrOH/ZnO heterostructures. Lower panel: Optimized structures of the vdW heterostructures ZrOH/ZnO (left) and the calculated corresponding
momentum-resolved surface density of states in the (1, 0, 0) surface (right).
\label{fig:sc}}
\end{figure}

Searching for appropriate substrate is crucial for the deposition of the ZrOH monolayer in experiments. As a 2D insulator with large band gap of 3.57 eV~\cite{43}, ZnO monolayer can match well with the ZrOH monolayer with a very small lattice mismatch ($\sim$ 0.9 $\%$). Fig.~\ref{fig:sc} shows that electronic states near the Fermi level are still dominated by the ZrOH films and the Dirac cone remains in the presence of the substrate. Due to substrate effects, a small gap is opened at the Dirac point and the inversion symmetry is broken. Therefore, the degenerate energy bands are separated in energy. The calculated momentum-resolved surface density of states in (1, 0, 0) shows that the edge states still exist at the BZ boundary, confirming that the topological phase of ZrOH monolayer is not affected (Fig.~\ref{fig:sc}).

\begin{figure}[t]
\centering
\includegraphics[width=0.85\linewidth]{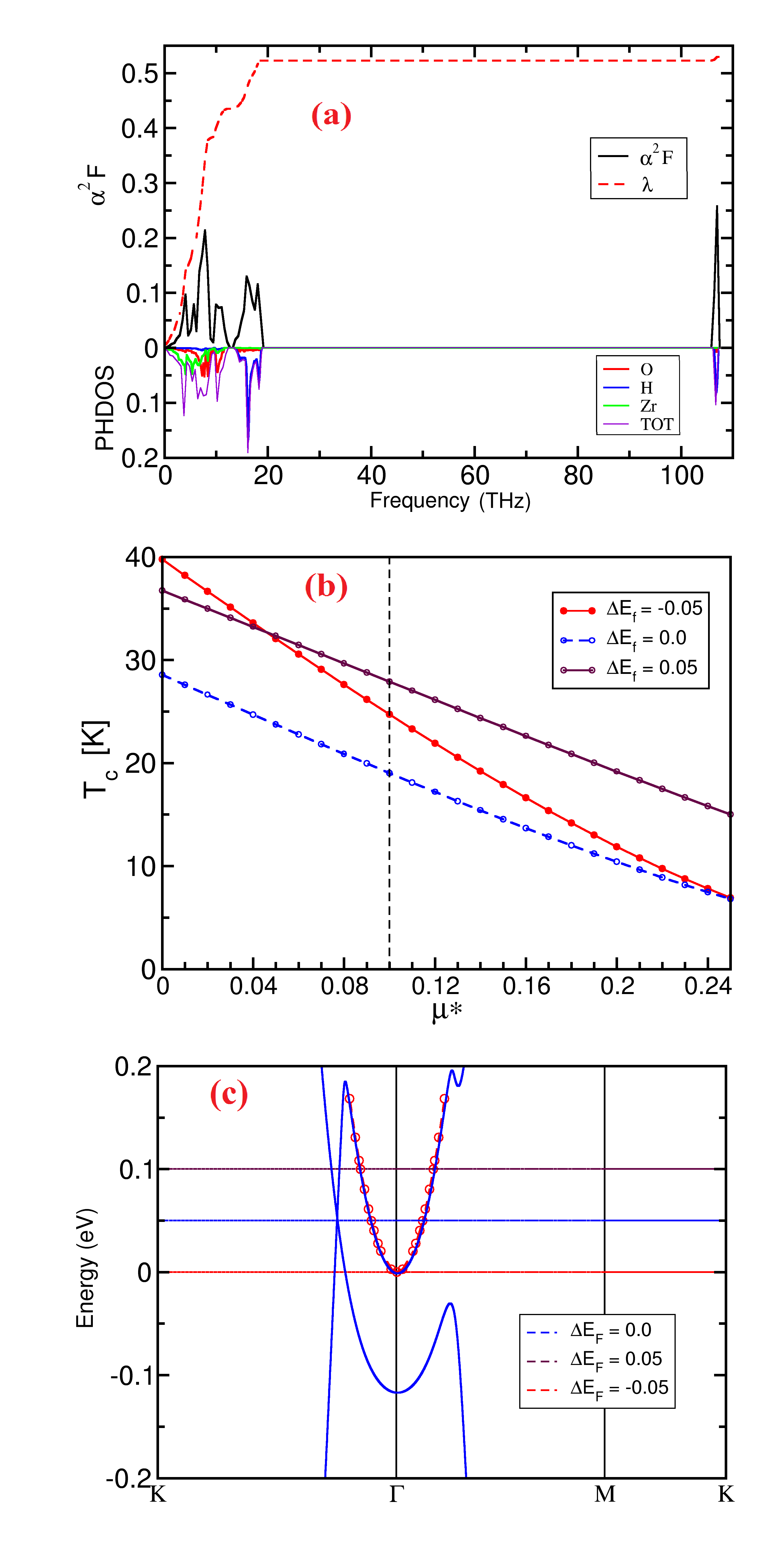}\\
\caption{(Color online) (a) The electron-phonon properties of ZrOH with the E$_{\rm F}$ set to 0.05 eV below the Dirac point (DP): the top part showing the Eliashberg spectral function $\mathrm \alpha^2F$ together with the integrated EPC constant $\lambda$, and the lower part showing the PHDOS. (b) The calculated values of $T_{\rm c}$ for different values of $\mu^*$ with the $\mathrm \alpha^2F$ set to the DP ($\Delta$E$_{\rm F}$= 0.0) and 0.05 eV above (below) it ($\Delta$E$_{\rm F}$ = 0.05⁡(−0.05)). The different trend of  $T_{\rm c}$ at ⁡$\Delta$E$_{\rm F}$ = −0.05 can be related to disparate $\lambda$. (c) Colorful dashed lines show the Fermi energy at the DP and above (below) it. Red circle shows that, $E$ = $2.64k^ 2 + 0.01k$ reproduces very well the \emph{ab-initio} band structure (solid blue line) around the Γ point.
\label{fig:form}}
\end{figure}

\begin{figure}
	\centering
	\includegraphics[width=0.99\linewidth]{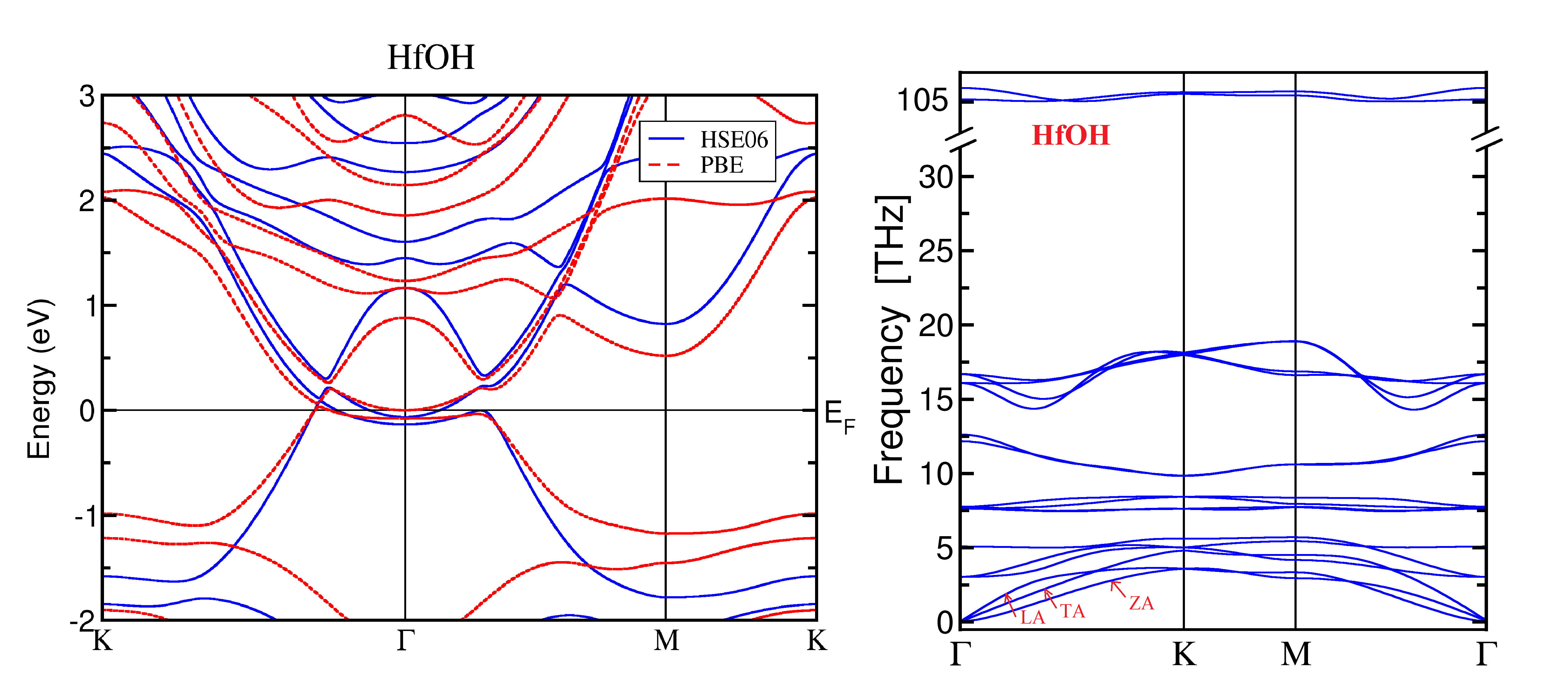}\\
	\caption{(Color online) Left panel: The bulk band structure for optimized HfOH using HSE06 and PBE. Right panel: The phonon dispersions for optimized HfOH nanosheets. LA and TA correspond, respectively, to the longitudinal and transverse waves with in-plane atomic displacements. ZA represents the transverse waves with out-of-plane atomic displacements. We calculate the slopes of in-plane acoustic branches in the vicinity of the $\Gamma$ point. The sound velocities in the $\Gamma$-K direction are obtained as 1.41 and 0.89 km/s for longitudinal and transverse atomic motions for HfOH, respectively.
		\label{fig:hubd}}
\end{figure}

  \renewcommand{\arraystretch}{1.1}
\begin{table}[h!]
	\caption{The calculated $T_{\rm c}$, $\lambda$ and carrier concentration ($n_{\rm s}$) at E$_{\rm F}$ for the Fermi level shift ($\Delta$E$_{\rm F}$)
		referring to the Dirac point (see Fig.~\ref{fig:form})}
	
	\begin{tabular}{c  c c  c} 
		\hline
		\hline
		$\Delta$E$_{\rm F}$ & $T_{\rm c}(K)$ & $n_{\rm s}$(cm$^{-2}$) & $\lambda$\\
		\hline
		
		-0.05 &  24.7 & 5.3$\times$10$^{11}$  & 0.52\\

		0.0 & 19.02 & 6.0$\times$10$^{13}$ & 0.72\\
		
		0.05 & 27.9 & 1.20$\times$10$^{14}$ & 0.71\\
		\hline
		\hline
	\end{tabular}
	\label{table:T}
\end{table}

Having calculated the band structure of the system and having analyzed its symmetries, we now explore possible superconducting behaviors of the system as well. Recent theoretical studies have revealed that the superconducting gap structure of Dirac semimetal can be interpreted from the orbital structure of Dirac semimetal and possible pair potentials~\cite{9, 10, 44}. In Dirac semimetal, bandcrossing points originates from bands (or orbitals) with different quantum numbers associated with a certain crystal symmetry. Therefore, the orbital mixing is prohibited in this symmetry-invariant momentum subspace including the Dirac point. However, if the Fermi surface does not coincide with such an invariant subspace, an orbital mixing is allowed on Fermi surface and Cooper pairing between electrons with different quantum numbers is possible~\cite{10, 44}.
In this case, he band structure of odd-parity superconductor might satisfy $PH_{0}(k)P=H_{0}(-k)$, therefore, it leads to $P{\Delta}_{i}(k)P=-{\Delta}_{i}(-k)$ Where ${\Delta}_{i}$ is the pairing potential~\cite{45} and P is an inversion operator acting on the orbitals within unit cell. In particular cases of ZrOH, two crossing bands are constructed from Zr-d and O-p orbitals. For such two orbital systems, the possible pair potentials can be classified into intraorbital (${\Delta}_{1}= {\Delta}\sigma_{0}s_{0}$ and ${\Delta}_{2}= {\Delta}\sigma_{z}s_{0}$) or interorbital (${\Delta}_{3}= {\Delta}\sigma_{y}s_{y}$, ${\Delta}_{4}= {\Delta}\sigma_{y}s_{x}$, ${\Delta}_{5}= {\Delta}\sigma_{x}s_{0}$ and ${\Delta}_{6}= {\Delta}\sigma_{y}s_{z}$)~\cite{10}. Here $s_{i}$ and $\sigma_{i}$ (i=0, x, y, z) are two Pauli matrices in the spin and orbital spaces respectively. 
In ZrOH monolayer, around the Dirac point (along the $\Gamma$-K), the two inverted bands have opposite parity and then inversion operator can be choose as $\hat{p}$ = $\sigma_{\rm z}$. Therefore, all interorbital pairings in ZrOH monolayer are odd under parity and one can naturally obtain an odd-parity pairing in ZrOH. Previous studies~\cite{44, 46} show that in the case of odd-parity superconductors, topological nature can be clarified by knowing the information of the Fermi surface. In other words, an odd-parity superconductor is a topological superconductor if the Fermi surface encloses an odd number of time-reversal-invariant momenta in the Brillouin zone.
As Fig.~\ref{fig:E} shows, the conduction band touches the Fermi level at the $\Gamma$ point and under small lattice compressive strain, this band crosses the Fermi energy. Since the Fermi surface in ZrOH encloses an odd number of TRIM in the Brillouin zone, any time-reversal-invariant odd-parity superconductivity realized on this Fermi surface is topological~\cite{44, 46}. As a result, ZrOH is an odd-parity topological SC with a gapless Majorana edge mode on the boundary. Moreover, the Zak phase at surface TRIM can be calculated from the products of the parity eigenvalues over the occupied states at the two corresponding bulk TRIM~\cite{47}. In particular, the Zak phase $\theta(\bar{\Gamma})$ at $\bar{\Gamma}$ is calculated from the product of the parity eigenvalues at the $\Gamma$ and the M points which gives $\theta(\bar{\Gamma})= \pi $ (mod $2\pi$). This nontrivial Zak phase, i.e., the parity inversion between the $\Gamma$ and M, means band inversion between the $\Gamma$ and M. As a result of this $\pi$ Zak phase, surface states should appear at the $\Gamma$ point. 

For the sake of completeness, we  also calculate the Zak phase by computing the 1D integral~\cite{47} 
 \begin{equation}\int_{0}^ {\left|G\right|} ~ \sum_{\rm i \in OCC}
<u_{{\bf k},i}|i\bigtriangledown_{{\bf k}}|u_{{\bf k},i}> d{\bf k}
\end{equation} 
where  G  is a reciprocal lattice vector perpendicular to BZ edge, $|u_{{\bf k},i}>$ is the periodic part of the lattice Block wave function with band index $i$ and momentum ${\bf k}$. This calculation shows again that the Zak phase is $\pi$ at ${\bar \Gamma}$ which is perfectly consistent with aforementioned result. 

As shown in Fig.~\ref{fig:form}, the vibrational modes are divided into three parts, with low frequencies dominated by Zr/O atoms, intermediate and high frequencies related mainly to H atoms. Comparing the $\mathrm \alpha^2F(\omega)$ with partial phonon density of states (PHDOS), it is clear that the most substantial contribution to the EPC comes from H and O modes at intermediate frequencies (Fig.~\ref{fig:form}). In our calculations we find the total EPC $\lambda$ to be 0.52 with major contribution from optical phonon mode around 10 THz. The superconducting $T_{\rm c}$ is estimated to be 24.7 K at the Fermi level which is located below the Dirac point. The Fermi level is 0.05 eV below the Dirac point which corresponds to hole concentration $n_{\rm s}$ =5.3$\times$10$^{11}$ cm$^{-2}$~\cite{48, 49} (see Appendix B). To consider the superconductivity around the Dirac point, we raise the Fermi energy and placing it at the Dirac point and above it (see Fig.~\ref{fig:form}).

By setting the Fermi energy at the Dirac point, the electron concentration becomes $n_{\rm s}$ =6$\times$10$^{13}$ cm$^{-2}$ which is driven by parabolic band. If we locate the Fermi energy at 0.05 eV above the Dirac point, both Dirac point and parabolic band take part in generating $n_{\rm s}$ =1.20$\times$10$^{14}$cm$^{-2}$ for the electron concentration at the Fermi level. The calculated results (Table~\ref{table:T}) show that the EPC $\lambda$ is increased to 0.7 by raising the Fermi energy while $T_{\rm c}$ is decreased by 23 \% at the Dirac point and then increased by 30 \% above band crossing. The value of the Coulomb pseudopotential $\mu^*$ is chosen to be 0.1 (dashed line in Fig.~\ref{fig:form}(b)) in these calculations. However, the calculated  $T_{\rm c}$ for several values of the Coulomb pseudopotential $\mu^*$ (Fig.~\ref{fig:form}) shows the strong impact of the electron-electron interaction on $T_{\rm c}$. The high EPC and superconductivity in ZrOH can be attributed to a Van Hove singularity of density of state at the Fermi level resulting from Zr-Zr distance reduction. These results are in good agreement with a previous report~\cite{23} which shows that the change in Zr-Zr distance is the most important factor in the appearance of superconductivity in ZrNCl.

\section{Summary and conclusions}
In summary, we have investigated the electronic structure and the EPC of ZrOH (HfOH). The calculated electronic and phonon properties of ZrOH show the coexistence of nontrivial band topology and superconductivity in this monolayer. In fact, the superconducting $T_{\rm c}$ is estimated to be 24.7 K. Our work demonstrates that 2D ZrOH can exhibit superconductivity and is a novel platform for realizing Majorana fermions. This work opens a new venue for future development in the search for two-dimensional topological superconducting materials and could be verified by experiments.

\section{Acknowledgments}

This work is supported by Iran Science Elites Federation.

\appendix

\section{Formation energy}\label{B}
The formation energies of surface groups for single-layer ZrX (X= OH, Cl, Br, F) are shown in Figure Fig.~\ref{fig:mass}, which are defined as
\be\label{dynamical_term}
E_{\rm f}=E_{\rm tot}(ZrX)-E_{\rm tot}(Zr)-E_{\rm tot}(X)
\ee

where $E_{\rm tot}$(ZrX) and $E_{\rm tot}$(Zr) stand for the total energies of ZrX and Zr monolayer, respectively, and $E_{\rm tot}$(X) is the total energy of O$_{\rm 2}$+H$_{\rm 2}$, $Cl_{\rm 2}$, $Br_{\rm 2}$ and $F_{\rm 2}$. Since the formation energy (-14.24) of ZrOH is larger than those of ZrX, probably its experimental production may be more feasible.

\section{ Carrier concentration}\label{C}
The carrier (hole) concentration, $n_{\rm s}$, at the Fermi level is related to the Dirac cone (Fig.~\ref{fig:sc}) with negligible part from a parabolic band around the zone center. Therefore, $n_{\rm s}$ is given by~\cite{50, 51}
\be\label{dynamical_term}
               \Delta E_{\rm F}=\hbar{\mid{v_{\rm F}}\mid}\sqrt{\pi n_{\rm s}}
               \ee

 where $\Delta E_{\rm F}$ is the Fermi level shift referring to the Dirac point and $v_{\rm F}$ = 0.5$\times$10$^8$ cm/s is the Fermi velocity. The Fermi level is 0.05 eV below the Dirac point which corresponds to hole concentration $n_{\rm s}$= 5.3$\times$10$^{11}$ cm$^{-2}$. To consider the carrier (hole) concentration around the Dirac point, we expand (see Fig.~\ref{fig:sc}) the electron dispersion, E(k), around the zone center to second order, E= ak$^2$ + bk , which provides a very good fit to the results of \emph{ab-initio} calculation with a = 2.645 and b = 0.011 (Fig.~\ref{fig:sc} (c)). By setting the Fermi energy at the Dirac point (Fig.~\ref{fig:sc}), the electron concentration becomes $n_{\rm s}$= 6$\times$10$^{13}$ cm$^{-2}$ which is driven by parabolic band. If we locate the Fermi energy at 0.05 eV above the Dirac point, both Dirac point and parabolic band take part in generating $n_{\rm s}$= 1.20$\times$10$^{14}$ cm$^{-2}$ for the electron concentration at the Fermi level.

\section{HfOH band structures }\label{Appendix:D}
The bulk band structure and phonon dispersions of optimized HfOH nanosheets are depicted in Fig~\ref {fig:hubd}. The longitudinal acoustic (LA) and transverse acoustic (TA) modes with in-plane atomic displacements are illustrated. ZA refers to the transverse waves with out-of-plane atomic displacements. The sound velocities in the $\Gamma$-K direction are obtained as 1.41 and 0.89 km/s for longitudinal and transverse atomic motions for HfOH, respectively.


\end{document}